# Tribologically induced crystal rotation kinematics revealed by electron backscatter diffraction

C. Haug[1,2], D. Molodov[3], P. Gumbsch[1,2,4], C. Greiner[1,2*]

[1] Institute for Applied Materials (IAM), Karlsruhe Institute of Technology (KIT), Kaiserstrasse 12, 76131 Karlsruhe, Germany
[2] KIT IAM-CMS MicroTribology Center (µTC), Strasse am Forum 5, 76131 Karlsruhe, Germany
[3] Institute of Physical Metallurgy and Materials Physics, RWTH Aachen University, Templergraben 55, 52056 Aachen, Germany
[4] Fraunhofer Institute for Mechanics of Materials IWM, Woehlerstrasse 11, 79108 Freiburg, Germany

## Abstract

Tribological loading of metals induces microstructural changes by dislocation-mediated plastic deformation. During continued sliding, combined shear and lattice rotation result in the formation of crystallographic textures which influence friction and wear at the sliding interface. In order to elucidate the fundamental lattice rotation kinematics involved in this process during the early stages of sliding, we conducted unlubricated, linear single pass sliding experiments on a copper bicrystal using sapphire spheres. Electron backscatter diffraction (EBSD) performed directly on the bulk surface of the wear tracks in the vicinity of the grain boundary reveals crystal lattice rotations by approximately up to 35°. Predominantly, the tribologically induced crystal rotations appear to be kinematically constrained to rotations around the transverse direction (TD) and occur in both grains, irrespective of load (2 to 8 N). We demonstrate that inverting the sliding direction (SD) inverts the sense of crystal rotation, but does not change the principal nature of rotation for the majority of indexed EBSD data. A lower proportion of the crystal lattice rotates much farther around TD (roughly up to 90°), accompanied by a superimposed crystal rotation around ±SD. Analysis reveals that sliding direction and grain orientation exert a systematic influence of how crystal rotations are accommodated. This is rationalized in terms of geometry, anisotropic wear track profiles and slip traces. Under specific conditions, combined crystal rotation and twinning are observed. These detailed insights into the fundamental nature of tribologically induced lattice rotation kinematics provide important guidance for applied research targeting materials with superior tribological properties.

[*] Corresponding Author: christian.greiner@kit.edu.

## Graphical Abstract

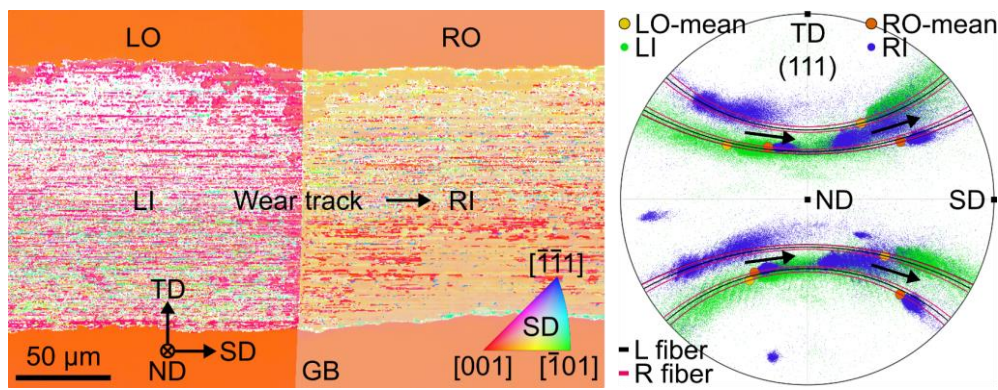

# 1 Introduction

Friction and wear are of central importance in tribological research – both historical and contemporary. This is largely because either parameter plays a key role in everyday engineering applications, providing manifold opportunities for energy conservation or the reduction of waste [1]. Friction and wear in turn are closely related to mechanical properties and modes of deformation of the materials comprising a sliding interface [2–4]. Facilitating such advances thus requires a detailed understanding of the fundamental mechanisms governing tribologically induced deformation in engineering materials.

The processes occurring in metal sliding systems strongly depend both on the materials in contact and the ambient conditions. Tribochemical reactions like oxidation are frequently observed in humid air or under lubricated conditions [5–7], while adhesion, material transfer and mechanical intermixing often dominate the evolution of metal-metal contacts [8–11]. During dry sliding of ductile metals, dislocation-mediated plasticity commonly alters the subsurface microstructure significantly, in turn influencing friction and wear. Alterations include grain refinement, texture formation, strain hardening or the formation of dislocation cell structures [3,12–17]. Complementary to the formation of new (sub-)grains, crystal lattice rotation, often around axes perpendicular to the sliding direction (frequently called transverse direction, TD), is a fundamental process involved in microstructure evolution of metals during dry sliding [18–20].

Crystal and grain rotation in general have been studied extensively, for example with regard to the interdependence of single and polycrystal deformation [21], influence of slip systems [22–25], texture formation [26,27], or the role of grain boundaries and interfaces [28–31]. In contrast to the aforementioned scenarios, knowledge about tribologically induced lattice rotation kinetics and kinematics remains rather limited. This is in no small part due to the fact that stress states induced by tribological loading are often very complex and can only be approximated employing severe simplifications – if at all (cf. e.g. [32]). Sphere-on-flat contact geometries constitute one of the rare examples for which an analytical model exists. However, the model by Hamilton remains limited to pure elasticity [33]. This makes computational modelling of tribologically induced plasticity and lattice rotations a very challenging task. In contrast, both are routinely accessible to suitable experimental approaches. One important aspect to consider in the present context is that tribologically induced deformation is usually constrained at the interface. Other than in unconstrained systems, arbitrary deformation may thus be accommodated by as little as three active slip systems [18,34]. By employing in-situ electron backscatter diffraction (EBSD), Chen et al. reported that lattice rotations of surface grains differ markedly from those further inside the material during tensile deformation of an AlMgSi alloy and may be associated with fewer active slip systems [35].

While most research on tribologically induced lattice rotation is primarily concerned with the later stages or steady-state of sliding [20,36–40], there is also some evidence of a similar rotation process occurring already after one single sliding pass [18,19,41]. The formation of one or multiple horizontal discontinuities (called dislocation trace lines) comprised of stationary edge dislocations was observed after a single sliding pass on polycrystalline copper and reported to be associated with a crystal rotation around TD [19,42,43]. A recent study combined cross-sectional high-resolution EBSD with crystal plasticity finite element modeling to analyze residual elastic deformation and lattice rotations induced by (effectively) single pass scratches with a Berkovich indenter on a copper (001) single crystal [44].

The possible dependence of crystal reorientation during sliding on crystal orientation has also received some limited attention. For instance, Lychagin et al. [45] recently studied surface reorientations on the side and worn faces of large, block-shaped (111) and (100) Hadfield steel single crystals in the steady-state sliding regime. They report a dependency of rotations on material flow due to wear lip formation, with the (111) single crystal exhibiting the most pronounced orientation changes [45]. Goddard et al. reported crystal rotation and twinning during the abrasion and scraping of various fcc metals, including copper and nickel with different crystal orientations, as well as discussing possible connections to friction [46].

Detailed insights regarding material and experimental parameters governing tribologically induced lattice rotation processes, their driving force and possible limitations, are however still lacking – especially in the early stages of sliding. Since Heilmann et al. reported that the degree of lattice rotation is likely greatest during the first of many sliding passes [18], it seems especially promising to further investigate early-stage lattice rotations.

Previous experimental works widely rely on analysis techniques like cross-sectional EBSD or transmission electron microscopy (TEM) methods for crystal orientation determination. While useful, both inherently limit the measurement area and require sophisticated, destructive specimen preparation routines – for example by focused ion beam (FIB), inevitably inviting the possibility of artifacts [47,48].

Here, we set out to simultaneously maximize the representativeness of the material volume probed while minimizing such artifacts by analyzing the near surface area directly using EBSD. That is without interfering with the sample's integrity in any way. In conjunction with a cluster-based evaluation of the crystal (re-)orientations measured, this approach allows us to quantitatively elucidate the impact of grain orientation, sliding direction and normal load on both the magnitude and fundamental mechanisms of tribologically induced lattice rotations.

## 2 Experimental

### 2.1 Sample preparation and experiments

Sapphire spheres (Saphirwerk AG, Switzerland) with a diameter of 5 mm were unidirectionally slid across a pure copper bicrystal specimen using a custom-built micro-tribometer (cf. **Fig. S1**) just once, producing wear tracks of 10 mm length. As depicted in the schematic in **Fig. 1a,** three wear tracks were positioned in parallel such that the bicrystal grain boundary (GB) was traversed perpendicularly after approximately 5 mm of sliding in each experiment. The sliding speed was 0.5 mm/s. For experiments 1 and 2, a normal load of 2 N was applied using deadweights. The sliding direction (SD) was inverted for experiment 2 by reverting the tribometer's travel direction. In experiment 3, a quadruple normal load of 8 N was applied. In conjunction, these experiments allow probing the influence of normal load, sliding direction and crystal orientation on crystal lattice rotation kinematics.

The frictional force at the interface of sapphire spheres and copper bicrystal was measured during the experiments using a piezo force sensor. The coefficients of friction (COF) reported in this work were calculated using $COF = F_F / F_N$, where $F_N$ denotes the constant normal load applied using deadweights as determined during normal load calibration, and $F_F$ the absolute value of the friction force counteracting displacement at the interface, irrespective of sliding direction.

All experiments were carried out at room temperature in a controlled dry nitrogen atmosphere. The sealed test chamber was flushed with $N_2$ (5N purity) for at least one hour before and during each test, yielding a relative humidity of roughly 0.1%. These conditions and materials were chosen to prevent tribo-oxidation and copper transfer to the sapphire sphere, while confining plastic deformation to the copper (cf. [5,42]).

A bicrystal of high purity (99.998%) copper was grown in a graphite mold in vacuum of $10^{-5}$ mbar by Bridgman method using specifically oriented single crystal seeds produced from the previously grown single crystal. The specimen with a diameter of 23 mm and a thickness of 1.8 mm was cut out from the grown bicrystal by electrical discharge machining. To ensure a low and homogenous initial dislocation density, it was ground, polished and electropolished before the experiments. The initial (undeformed) orientation of both grains was determined by EBSD in the vicinity of experiment 1 (cf. section 2.2 for details). As can be inferred from the (111) and (100) pole figures in **Fig. 1b/c**, the mean orientation of the left grain (yellow dots) is close to a cube orientation, with the <100> direction only ~1.5° from the surface normal direction (ND). In the right grain (orange dots), the <100> direction is ~17° from ND. The crystallographic directions aligned with SD / ND / TD in the left and right grain are approximately $[\bar{6}\ 1\ 0]$ / $[0\ 0\ \bar{1}]$ / $[\bar{1}\ \bar{6}\ 0]$ and $[\overline{12}\ \bar{1}\ 3]$ / $[\bar{2}\ 1\ \bar{7}]$ / $[0\ \bar{6}\ \bar{1}]$[1]. With a misorientation of ~21.5°, the grain boundary is a common high-angle grain boundary, typical for many polycrystalline materials.

## 2.2   Sample characterization and data analysis

In order to facilitate an assessment of the lattice rotations induced by tribological loading in each grain, the crystal orientations of the copper lattice were measured using EBSD at the intersection of the grain boundary and each wear track. To this end, a Bruker Quantax eFlash HD detector (Bruker Nano Analytics, Germany) was employed, mounted on a FEI Helios NanoLab 650 dual beam scanning electron microscope (SEM) / FIB (FEI Company, USA). SEM operating conditions were 20 kV acceleration voltage and 13 nA probe current, with a step size of 0.45 µm (experiments 1 and 2) and 0.9 µm (experiment 3). The EBSD detector was operated using a binned detector resolution of 160x120 $px^2$ and an exposure time of 7 ms. Standard Hough-based indexing (see for example [49–51]) was performed by the Bruker acquisition software Esprit 2.2. The threshold values for accepting an individual measurement point (also referred to as data point or pixel) as indexed were set to a minimum of six indexed Kikuchi bands with a maximum mean angular deviation (MAD) of 2°. Any pixels not meeting these criteria were treated as non-indexed.

Orientation data was exported for further analysis and visualization in MATLAB using the free toolbox MTEX [52]. To obtain a consistent rectangular area in all EBSD map plots, measurements were slightly cropped at the edges for visualization, but all available orientation data was evaluated for calculations, pole figure (PF) and inverse pole figure (IPF) plots. Subdivision of each EBSD data set into regions and clusters based on spatial position or crystal orientation of data points was achieved by combining MTEX's automatic grain segmentation algorithm (threshold angle 1.25°) with manually defined polygons.

---

[1] The Miller indices for the right grain are not orthonormal due to rounding. More precise values with deviation from orthogonality between SD / ND / TD << 0.1°: [-3.494 -0.262 0.889] / [-0.918 0.504 -3.460] / [0.126 -3.570 -0.554].

Supplementary characterization of wear track topography and sample surface roughness was performed using a Sensofar Plu neox optical 3D profilometer (Sensofar, Spain), operated in interferometric vertical shift image mode. The mean areal surface roughness $S_A$ was determined from seven measurements in the left and right grain each, yielding mean values of 63 nm (standard deviation: 16 nm) and 61 nm (standard deviation: 10 nm), respectively (62 nm with standard deviation 13 nm in total). Mean wear track profiles were obtained by evaluating one measurement at the intersection of wear track and grain boundary per experiment using custom MATLAB code.

## 3   Results

### 3.1   Crystallographic orientations measured by EBSD

In order to assess the lattice rotations induced by each experiment, we performed bulk EBSD on the bicrystal sample surface as is. **Fig. 2a** displays the EBSD measurement acquired in the GB's vicinity for experiment 1 (cf. raw data available at [53]). The map is colored with respect to SD and covers an area of about 320x215 µm². Since its height exceeds the wear track width (~145 µm), it can intuitively be segmented into four distinct regions: One outside of the wear track for each grain (left outside, LO and right outside, RO, comprised of two parts each – above and below the wear track), as well as a segment inside the tribologically deformed wear track per grain (left inside, LI and right inside, RI). Thus, LO and RO represent the two grains' initial crystal orientation, while LI and RI can be evaluated to assess the tribologically induced lattice rotations in each grain.

An analogous separation into four respective regions may be performed for experiments 2 (sliding direction inverted) and 3 (quadruple load) depicted in **Fig. 2c/e**. Comparing the color associated with regions LO and RO for all three experiments confirms that the crystal orientation of the bicrystal grains with respect to SD is very similar across experiments. **Fig. S2** corroborates that the same is true for the crystal orientations with respect to ND and TD. It may also be referred to for appraising the crystal rotations inside the wear track (LI / RI) with respect to these directions.

In all EBSD maps presented, white pixels constitute non-indexed data points. **Table 1** provides an overview of the indexing statistics (percentage of data points indexed) and data quality across measurements and regions. Here, band contrast (BC, a qualitative measure of Kikuchi bands' contrast with respect to the pattern background) and the number of indexed Kikuchi bands (out of a maximum of twelve total for face-centered-cubic (fcc) crystal symmetry) are assessed as measure for data quality. In conjunction, **Table 1** and **Fig. 2** allow for five observations: All data points are indexed in regions LO and RO outside the wear track (as a result of our definition of these regions, which here excludes non-indexed pixels), whereas a substantial proportion of unindexed pixels are present inside the wear track (LI / RI): Around 10-35% for experiments 1 and 2, and up to 70-90% in experiment 3. A systematic trend for a larger fraction of points to be indexed successfully in the left grain (LI) than the right grain (RI) inside the wear track is apparent across all experiments, by approximately 5-25 percentage points. This may be due to differences in crystal orientation (after crystal rotation), as some crystal orientations produce diffraction patterns that are more difficult to index for the software, for example if a zone axis is located near the pattern center. Outside the wear track (LO and RO), the mean number of indexed band is around 0.6 - 1.2 bands larger in the right grain than in the left grain. Considering all data (i.e. all regions), it is evident that indexing is much better in experiments 1 and 2 than in experiment 3. Average data quality is systematically worse for experiments and / or regions with poor indexing, both in terms of a lower BC and fewer indexed bands.

In order to assess the lattice rotations occurring as a result of tribological loading in each grain, the crystal orientations found inside the wear track after deformation need to be compared to the original crystal orientations outside of the wear track in the respective grain, representing the undeformed state. Corresponding (111) pole figures visualizing these lattice rotations for experiments 1-3 are presented in **Fig. 2b/d/f**, next to the corresponding EBSD maps (**Fig. 2a/c/e**). Each small dot in the pole figures represents one indexed pixel in the corresponding EBSD data set. In each pole figure, all data points in each of the corresponding data set's four regions share the same color: The initial grain orientations (LO and RO) are colored in yellow and orange, those inside the wear track in the respective grains (LI and RI) in green and blue. As the data points comprising each region were successively plotted on top of each other (in the order LI, RI, LO, RO), small portions of regions sharing similar crystal orientations may be partially obscured by data from a subsequent region, plotted on top.

In all three experiments, orientations in LO and RO are each tightly clustered, as may be expected from the crystal lattice in a low defect density grain. In contrast, each of the four {111} poles for regions LI and RI appear spread out from the corresponding initial orientation along distinct curved lines (see exemplary arrows in **Fig. 2b/d/f**). This clearly indicates substantial lattice rotation inside the wear track in both grains. While this spread of orientations extends in SD direction for experiments 1 and 3 (in **Fig. 2b/f**), its sense of direction is opposite (–SD) for experiment 2 (**Fig. 2d**), in which the sliding direction was inverted. This clearly demonstrates that changing the sliding direction impacts the observed lattice rotations. Increasing the load fourfold from experiment 1 to 3, however, does not qualitatively change the crystal orientations observed, with similar features in both pole figures. The fact that fewer data points are visible in **Fig. 2f** in comparison to **Fig. 2b** is due to poorer indexing (cf. **Table 1**). It is of note that a gap without any data points is present between both initial (LO / RO) and deformed (LI / RI) regions at some of the {111} poles in **Fig. 2b/d/f**, as for example marked in **Fig. 2b**. In order to gain a more detailed understanding of tribologically induced lattice rotation kinematics, it is necessary to scrutinize the lattice rotations observed inside the wear track (regions LI and RI) for the three experiments in greater detail.

## 3.2 Segmentation of orientation data inside wear track into clusters

The spreading of the crystal orientations in regions LI and RI along curved lines in **Fig. 2b/d/f** (from the initial orientations LO and RO) requires further consideration. Since pole figures are two-dimensional, stereographic projections of crystallographic poles in three-dimensional space, the projected trace of a plane parallel to the coordinate axes SD and ND is observed as a curved line in the pole figure. It is an inherent property of any such plane that it is also perpendicular to the transverse direction TD. The physical meaning of the {111} poles in **Fig. 2b/d/f** following such curved lines is thus that the corresponding copper lattice has performed a rotation within a plane perpendicular to TD, i.e. around a rotational axis parallel to TD. This rationalization is illustrated in **Fig. 3a/b** for the left and right grain of experiment 1, respectively. The black and purple lines represent a fiber of all theoretically possible orientations obtained for an *ideal* rotation of the mean initial orientations LO-mean and RO-mean (yellow / orange dot) *exactly* around TD (by all possible angles). The eight curved lines in each pole figure thus depict the path of the four {111} poles when performing a 360° rotation around TD, starting at LO-mean (**Fig. 3a**) or RO-mean (**Fig. 3b**). At the points where the fibers intersect with the circumference of the pole figure, the corresponding poles cross from the upper to the lower hemisphere or vice-versa. As all poles are projected to the upper hemisphere for plotting, the fibers consequentially appear reflected at the pole figure center (antipodal symmetry). Two pairs of black and purple arrows (with one and two stars) illustrate this for two {111} poles in the left and right grain, respectively. The corresponding data points can be seen traversing hemispheres along these portions of the fibers.

This delineation allows further segmenting the orientation data within regions LI (**Fig. 3a**) and RI (**Fig 3b**) into clusters of similar orientation. For the left grain, all orientation data inside the wear track in **Fig. 2a** (LI, green dots) was replotted in **Fig. 3a**. In conjunction with the fiber, one can clearly discriminate specific clusters of orientation data. The data points on the fiber and close to LO-mean constitute a first cluster CL1, encircled by the red polygon and colored in blue for all four {111} poles. Their misorientation with respect to LO-mean is characterized by a rotation in the range of approximately $5 - 35°$ almost perfectly around TD.

A second cluster CL2 (enclosed by the second red polygon and colored in purple) is slightly shifted towards TD in the pole figure. It is located at rotational angles around TD of roughly $55 - 90°$ with respect to LO-mean, but cannot be described by a pure rotation around TD because of the shift of data points in TD direction away from the fiber (see arrows in **Fig. 3a**). All other data points inside the wear track that are not part of clusters CL1 or CL2 remain colored in green.

**Fig. 3c** shows the clusters with the same coloring in an IPF with respect to ND. Data points in CL1 appear concentrated near the bisector in the vicinity of the <001> corner (see arrow). It is important to consider how many data points in LI belong to either cluster. This information is provided in a separate, next section, which is dedicated to the quantification and data quality of all clusters across grains and experiments.

Similarly, the orientation data inside the right grain (RI, blue) can be discriminated into three distinct clusters (**Fig. 3b**). Cluster CR1 (green) is comprised of data points that have rotated almost perfectly around TD by up to 20° from RO-mean. Akin to CL2 in the left grain, cluster CR2 (yellow) consists of data points at higher rotational angles around TD (~40 – 65°) that are slightly displaced in TD direction and can thus not be described by a rotation only around TD. A third cluster CR3 (colored in brown) clearly stands out. For a single {111} pole, it coincides with CR1 but is otherwise located at completely different locations in the pole figure, far away from the fiber. All other data points remain blue. In the IPF in **Fig. 3d** all three clusters form distinct clouds of points. CR2 data points are located closer towards the bisector compared to CR1 (see arrows). **Fig. S3** presents (110) and (100) pole figures of the same clusters for both grains. It demonstrates that the present discrimination of orientation data into clusters remains valid also when considering other poles than (111).

In order to assess the influence of sliding direction and normal load on crystal rotations inside the wear track, an analogous segmentation of the corresponding EBSD data into clusters was performed for experiments 2 and 3. As the pole figures in **Fig. 4a/b** reveal, clusters featuring similar characteristics to those in experiment 1 may be identified for experiment 2 (inverse sliding direction, in –SD). In contrast to experiment 1, however, crystal rotations now follow the fiber in the opposite direction (see horizontal arrows, CL1 / CR1). This means that these data points are associated with a rotation around –TD (the inverse of TD). Similarly, data points in CL2 and CR2 are now systematically displaced in –TD, i.e. in the opposite direction when compared to CL2 / CR2 in experiment 1 (**Fig. 3a/b**). An equivalent to CR3 in experiment 1 cannot be identified for the right grain for experiment 2. As detailed in section 3.1, experiment 3 (quadruple load, see **Fig. 4c/d**) qualitatively exhibits a behavior similar to that in experiment 1. Displacement in TD of data points inside CR2 is less pronounced in experiment 3. Additionally, a cluster similar to CL2 in the left grain cannot clearly be localized due to a lack of data points. In contrast, clusters CR1-CR3 are found at crystal orientations almost identical to those of experiment 1. CR2 however comprises a smaller area than in experiment 1. The changes in crystal directions induced by lattice rotation with respect to ND for experiments 2 and 3 are summarized in **Fig. S4**. Notably, the mean Miller indices of the crystal directions aligned with ND in CR1 are very different between experiment 1 (**Fig. 3d**) and experiment 2 (**Fig. S4b**). This means that when inverting the sliding direction, different crystal directions are aligned with the tribological interface's normal direction after tribological loading. A general trend of {111} slip plane normal directions to align with ND, as some previous works regarding tribologically induced texture formation suggest [15,37], is not observed.

### 3.3 Cluster quantification and data quality

Each of the three types of cluster (CL1 & CR1 / CL2 & CR2 / CR3) is associated with a characteristic type of lattice rotation mechanism, as will be discussed in section 4.1. In order to be able to assess their mutual significance, it is therefore important to consider which fraction of the data points in each grain belong to each type of cluster, as well as the EBSD data quality that each cluster is associated with. Since the pole figures in **Figs. 3/4** are comprised of tens of thousands of points, plotted in close vicinity and on top and of each other, they are not well suited for appraising the number of points per cluster. Consequentially, **Table 2** provides a concise overview of the proportions of the number of data points that each cluster in each experiment is comprised of (IP, indexed points) compared to the other clusters per grain. It is important to remember that as per definition, clusters exclusively consist of indexed data points: An indexed data point belongs to a given cluster if any of its {111} poles lie within the polygon defined on the unit sphere in terms of the corresponding (111) pole figure, cf. **Figs. 3/4**. As a means to judge the prevalence of each type of cluster, the absolute number of points that each cluster is comprised of itself is not very useful. Instead, it is necessary to relate the number of indexed pixels (IP) in each cluster with the number of points that the whole region inside the wear track of the corresponding grain is made up of (parent region, either LI or RI of the corresponding experiment). Columns PI (proportion of *indexed* pixels) and PT (proportion of *total* pixels) in **Table 2** supply this ratio: The percentage that the indexed pixels in each cluster (IP) constitute in terms of the *indexed* and *total* pixels of its parent region. Column PI demonstrates that cluster type 1 (CL1 / CR1) is by far the most common, comprising the majority of around 60 – 80% of the indexed pixels across all experiments and in both grains. Type 2 clusters (CL2 / CR2) are generally less prevalent, including only around 10 – 30% of indexed data points. Cluster type 3, only observed in experiments 1 and 3 in the right grain (CR3), constitutes only a very small fraction of data points (around 2%).

It is reasonable to pose the question how distorted this assessment may be by non-indexed pixels in the EBSD measurements. As column PT in **Table 2** shows, around 40-50% of *all* pixels still belong to cluster type 1 when including non-indexed pixels for experiments 1 and 2. The relative proportion of cluster types naturally does not change between PI and PT, with orientations belonging to cluster type 1 roughly around 2-7 times more frequently than to cluster type 2.

**Table 2** also allows ascertaining the data quality that is associated with each type of cluster, similar to the results presented per region in **Table 1** (section 3.1): It is evident that across all experiments and in both grains, type 1 clusters systematically possess the highest data quality of all clusters, both in terms of a high band contrast and a large number of indexed bands. Type 2 clusters, associated with larger rotational angles around TD and a displacement of poles in ±TD in the pole figures in **Figs. 3/4,** generally exhibit lower data quality. Furthermore, data quality is significantly worse for experiment 3 across all clusters compared to experiments 1 and 2.

### 3.4 Friction and wear track topography

While the detailed analysis of tribologically induced crystal lattice rotations by means of EBSD is at the heart of this work, friction and wear track topography are also important. Briefly considering these more macroscopic system responses of the tribological experiments aids the discussion of crystal lattice rotations and plastic deformation by providing context and a complementary, broader perspective.

**Fig. 5** depicts the COF during experiments 1 and 3 (black and purple graphs in **a**) and experiment 2 as a function of sliding distance. The grain boundary was traversed after approximately 5 mm of sliding in each experiment (vertical line). In experiments 1 and 3, mean COF values are systematically higher in the left grain (~0.32). than in the right grain (~0.30), with a step of approx. 0.02 in COF at the GB. While the absolute difference in mean COF per grain is very small (~0.02), the graphs for experiments 1 and 3 agree extremely well, especially considering the inherent spread and substantial uncertainty typical for tribological measurements, even in spite of careful calibration (cf. e.g. [54]). Experiment 2 with inverted sliding direction (**Fig. 5b**) shows a monotonic increase in COF and no step when traversing the GB but the mean COF in the left grain (~0.37) also is higher than in the right grain (~0.33). Absolute values are slightly elevated by approx. 0.03-0.05 in experiment 2 compared to experiments 1/3.

The topography of the wear track also contains useful information concerning the tribologically induced plastic deformation. **Fig. 6a** shows the mean wear track profile in the left (yellow, L) and right (orange, R) grains for experiment 1, viewed along SD. These mean profiles were extracted from the 3D profilometry measurement in **Fig. 6b** by computing a profile in the ND-TD plane at every pixel along SD. The GB is located approximately at the dotted vertical line. Averaging all profiles left and right of the GB yields the mean profile for the left and right grain as depicted in **Fig. 6a**. Before profile extraction, the height map in **Fig. 6b** was corrected for misalignment by subtracting an equalization plane, fitted to the undeformed portions of the map. **Fig. 6a** shows that the pile-up at the left and right edges of the wear track is anisotropic in the two grains. While it is almost identical at the left edge of the wear track (in TD direction) for both grains, it is almost twice as high in the left grain at the right edge of the wear track (in – TD direction). Fig. **6c/d** present profiles obtained in a similar fashion for experiments 2 and 3. While the mean profiles for experiment 3 (**Fig. 6d**) qualitatively exhibit the same features as experiment 1 (albeit the absolute profile heights are almost three times as large), relative pile-up heights in the left and right grains are reversed for experiment 2 (**Fig. 6c**) when compared to experiment 1: In the case of experiment 2, pile-up is approximately one and a half times as high in the right grain as in the left grain at the right edge of the wear track. Similar to friction, these findings indicate that plastic flow in both grains is anisotropic with respect to sliding direction and grain orientation.

# 4    Discussion

## 4.1    Overview: Three fundamental types of crystal lattice rotation

As detailed in section 3.2, a majority of the EBSD data points inside the wear track may be pooled into clusters sharing a similar, characteristic kind of crystal orientation with respect to the corresponding grain's initial orientation (**Figs. 3-4**). The purpose of this section is to establish the three types of crystal lattice rotations associated with each of the three types of cluster. As demonstrated, this segmentation is applicable to all three experiments, albeit not every kind of cluster exists for each experiment and grain.

Crystal rotation type 1 comprises clusters CL1 and CR1 and may be categorized as the crystal lattice having performed a rotation around axes very close to the transverse direction (+TD) or its inverse (-TD), i.e. in good agreement with the orientation fibers in **Figs. 3/4**. Rotational angles are comparably small, roughly up to 35°, depending on the sliding direction and grain. It has to be pointed out that this range of rotational angles is of course subject to the (manual) definition of the clusters – which is not adamant, especially for the left grain, with data points existing almost continuously along the fiber (cf. **Fig. 3a** and **Fig. 4a** as well as section 4.2 regarding the density of data points along the fiber in the left grain). The rotational angles should thus be understood as an order of magnitude rather than a set of precise values. The same restriction applies to all values previously presented as well as the forthcoming discussion of quantitative parameters associated with clusters and types of rotation. The concept of lattice rotations (nearly) around TD is what is the defining characteristic for type 1 rotations, inherently limiting the maximum rotation angles that may be appraised as a cutoff.

In contrast, type 2 lattice rotations (clusters CL2 / CR2) are characterized by the lattice having performed a rotation around TD with larger rotation angles (very roughly in the range of 20 – 90°) and, most importantly, systematically deviating from an ideal rotation around TD. This deviation is visible from the displacement in either +TD (experiment 1, **Fig. 3a/b**) or –TD (experiment 2, **Fig. 4a/b**) direction, away from the corresponding orientation fiber in the pole figures. This displacement may be rationalized as an additional lattice rotation around –SD (experiment 1) or +SD (experiment 2), superimposed on the rotation around ±TD. Of course, the exact lattice rotation axes may also contain an ND component (e.g. for CR2 in experiment 3) but this appears to be small. It is important to stress that when changing the sliding direction between experiments 1 and 2, the sense of rotation interchanges simultaneously for both the rotations around TD and SD.

Finally, a third type of characteristic lattice rotation (type 3) is defined by clusters CR3, located completely secluded from the fiber, save for one of the four poles (cf. **Fig. 3b** and **Fig. 4d**). Type 3 is found exclusively in the right grain for experiments 1 and 3. As CR3 is comprised of four clearly demarcated clouds of poles, calculating a mean misorientation between RO and CR3 is feasible and may help shedding light on the deformation mechanism it is associated with. This misorientation is 57.0° around $[\overline{8}\ \overline{9}\ 11]$ for experiment 1 and 56.6° around $[\overline{4}\ \overline{5}\ 6]$ for experiment 3. Both are close to a face-centered-cubic Σ3 twinning misorientation of 60°<111> (cf. e.g. [55]). If one considers the coincidence of one CR3 pole with CR1, it seems sensible to also calculate the mean misorientation between CR1 and CR3, which is 59.6° around $[\overline{11}\ \overline{12}\ 12]$ and 58.8° around $[\overline{7}\ \overline{8}\ 8]$ for experiments 1 and 3, respectively. Both values are even closer to 60° around <111>. This strongly suggests that the crystal lattice in CR3 has performed a type 1 crystal rotation followed by deformation twinning. Whether these two processes happen concurrently or successively cannot be ascertained definitively. Furthermore, this combination of rotation and twinning implies that it is technically incorrect to speak of type 3 lattice rotations. For the sake of simplicity and consistency, this wording is retained anyways throughout this work.

Deformation twinning is frequently observed in face-centered cubic crystals, especially those with low stacking fault energy (SFE) [55]. It is thus conclusive that twinning is not a primary mode of deformation in the present experiments with medium SFE copper. Furthermore, the fact that type 3 rotation and twinning is only observed in the right grain, and only when maintaining the same sliding direction (experiment 1 and 3) may be rationalized considering the mechanisms of twin formation [56]. In their review, Beyerlein et al. point out that the propensity for deformation twinning in face-centered-cubic crystals depends on the resolved stresses acting on twin systems and thus the sense of shear (and implicitly, sliding direction) [57]. The orientation relationship between CR1 and CR3 for both experiments 1 and 3 is compatible with activation of the twin system $(\bar{1}\ \bar{1}\ 1)[2\ \bar{1}\ 1]$ (cf. e.g. [55]).

## 4.2   Relative significance of types of lattice rotation and influence of data quality

Having established the three fundamental types of lattice rotation, their relative significance is of special importance: Which type is most commonly observed, and which influence on the relative proportions does EBSD data quality possess? As detailed in **Table 2** in section 3.2, type 1 lattice rotations are observed 2-7 times more frequently than type 2 lattice rotations and possess the best data quality. Type 3, combined rotation and twinning, are only observed for a low fraction of data points (around 2%), and only for experiment 1 and 3 in the right grain. This indicates that type 3 may only occur under more specific conditions than type 1 and 2 – possibly a combination of a specific, local stress state and crystal orientation. While the magnitude of rotational angles is discussed in more detail in the next section, **Table 2** demonstrates data quality is generally higher in clusters exhibiting smaller rotational angles (CL1 / CR1, type 1) as opposed to higher rotational angles (CL2 / CR2, type 2). In conjunction with a systematically lower data quality for experiment 3 with quadruple load, this may be rationalized by considering that higher rotational angles and higher load are naturally associated with more plastic deformation and, more specifically, a higher dislocation density. This inherently leads to increased distortion of Kikuchi patterns and worse indexing. While one may thus suspect that non-indexed pixels could be associated with even more deformation (and thus, hypothetically, crystal lattice rotation), this cannot be ascertained. Restricting the EBSD data to pixels possessing at least 10 indexed bands before carrying out the present analysis systematically increases prevalence of type 1 over type 2 lattice rotations. Similarly, experiment 3 exhibits the highest type 1 PI values (~70-80%, cf. **Table 2**) of all experiments and rotations around SD of type 2 data points are smaller than for experiment 1. In conclusion, this means that the (best) data available strongly suggests greatest generality of type 1 lattice rotations, followed by types 2 and 3. In other words, this implies that a rotation around ±TD is the predominant mechanism at the heart of tribologically induced lattice rotation kinematics. The information presented on data quality is also of interest with regard to the novel approach of using EBSD directly on the tribologically loaded surface to determine tribologically induced crystal rotations on a large scale while obviating the risk of artifacts: One has to expect a deterioration of data quality and indexing proportions in situations with further increased plastic deformation, e.g. when using even higher normal loads than 8 N or multiple sliding passes. While extremely useful to study tribologically induced lattice rotations for cases with low to medium plastic deformation and dislocation densities (such as the experiments discussed), the present approach may be limited in such situations.

In order to further validate the reliability of our approach in the case of experiment 3 with comparably poorer EBSD indexing rate, we extracted a TEM foil parallel to SD in the center of the wear track at the GB by FIB milling (cf. e.g. [47]), and measured the crystallographic orientations using on-axis Transmission Kikuchi Diffraction (TKD) [58]. The TKD orientation maps and (111) pole figure evaluation in **Fig. S5** in the supplementary information (described and assessed there in detail) clearly corroborate that the experiment 3 EBSD data as depicted in **Fig. 4c/d** – even despite relatively poor indexing – appropriately captures the core mechanisms of crystal rotation kinematics: type 1 crystal rotations around TD in both grains, type 2 rotations in the right grain (rotations around TD, superimposed with components around SD and -ND), and combined type 1 rotations and twinning (type 3) in the right grain. Furthermore, an orientation gradient in the rotated subsurface area parallel to the sliding direction next to the GB in the right grain in **Fig. S5a** indicates that the GB apparently exerts an influence on lattice rotations in its immediate, very close proximity. In-depth discussion of these results obtained by a destructive technique is beside the point in the present work, but considering the differences in measurement area, the possibility of artifacts due to TEM foil preparation as well as the GB's influence, agreement between EBSD and TKD is astonishingly good – highlighting and substantiating the applicability and advantages of the present EBSD approach. This also demonstrates that the present procedure of using large-scale EBSD serves a means to exclude possible effects of the GB's presence on the study of lattice rotation kinematics.

Returning to the matter of judging the relative significance of lattice rotation types based on EBSD data points per cluster, one may argue that considering the absolute number of data points per type of cluster may distort this assessment due to the differences in cluster area. However, scrutinizing the surface density of data points per cluster area (fraction of the unit sphere surface area enclosed by the polygon defining each cluster) for experiment 1 substantiates the conclusions presented (cf. **Table S1**): In the left grain of experiment 1, this surface density steadily declines from CL1 over the area between CL1 and CL2 to CL2 (cf. **Fig. 3a**). In the right grain, the same trend applies, with a local minimum in between CR1 and CR2, where almost not data points are observed (cf. **Fig. 3b**). These relative proportions of density clearly demonstrate that for the most significant portion of data points, the crystal lattice performs type 1 rotations around ±TD. Only for smaller, less significant fraction of data points, does it rotate much farther, and then systematically performs a superimposed rotation around ±SD (type 2).

Considering this assessment in conjunction with literature [18–20,36,41] further corroborates the following hypothesis: The fundamental character of subsurface crystal lattice rotation around ±TD (type 1, but also as a part of types 2-3) induced by dry sliding on copper is a consequence of the kinematic constraints imposed onto the sliding interface by tribological loading [34], and not a consequence of grain orientation, sliding direction or load at the interface.

The extent to which there is a systematic influence of slip system alignment (with respect to the normal and shear loading directions ND and SD) on how crystal rotation is accommodated (especially for type 2 lattice rotations), is discussed in detail in section 4.4. The fact that lattice rotations around ±TD were previously observed for a variety of material combinations and contact geometries such as ball-on-flat [19,41], block-on-ring [18] and flat-on-flat [20,36]) suggests further generality: It seems likely that the rotation kinematics are not a consequence of specific stress states characteristic for different contact geometries (e.g. the stress field below a moving sphere [33]) but rather the combination of normal and shear loading itself at a partially constrained interface. Expanding on this, one may wonder about situations in which crystal rotations are additionally constrained, for example in small-grained material states. While it seems likely that the general situation does not change, and type 1 crystal rotations are still prevalent, one may speculate that due to the additional constraints imposed by the GBs and different crystal orientations and slip system geometry in adjacent grains, type 2 crystal rotations increasingly deviating from perfect rotations around ±TD could become more important. The same reasoning could be applied to situations with even higher loads than in experiment 3, in which subgrain formation has been shown to ensue as a consequence of load [41]. As the GB does not exert a discernible influence on lattice rotations observed by EBSD in the present case (see next section), this is probably only relevant for grain sizes in the range of a few microns or less.

## 4.3 Spatial distribution, magnitude of rotational angles and physical rationalization of types of lattice rotation

It is of special interest whether type 1-3 lattice rotations are associated with certain characteristic positions within the wear track. As an example, one might suspect that lattice rotations are especially constrained in close proximity to the grain boundary (cf. **Fig. S5**), resulting in smaller lattice rotation angles. Similarly, the lateral position of a data point in the wear track (i.e. the distance from the wear track edges) could be relevant. **Fig. 7** depicts the spatial distribution of EBSD data points of experiment 1 for each cluster (sliding direction of the sapphire sphere from left to right, in SD, as marked in each image). Each part of the figure depicts an EBSD map plot colored with respect to SD (similar to **Fig. 2a**), exclusively showing the indexed data points belonging to a single cluster. All other indexed and non-indexed data points are omitted from the plot. In conjunction with the white lines demarcating the wear track edges and the GB's location, this allows appraising the spatial distribution of crystal lattice rotation types within each grain at a glance. In both grains, type 1 (CL 1 / CR1) as well as type 2 (CL2 / CR2) lattice rotations appear to be rather evenly distributed across the whole corresponding region LI / RI. It is of special note that type 3 (CR3) data points also appear distributed over all of RI in the right grain. This suggests that combined lattice rotation and twinning is a fundamental mechanism and not the byproduct of a localized event. In conclusion, no clear correlation between position and magnitude of lattice rotation can be discerned for experiment 1 on the scale of the present experiment. If anything, one may note a slight over-representation of clusters CL1 and CR1 at the sides of the wear track. The situation may of course be different in very close proximity (a few microns, cf. **Fig. S5**) to the GB but such difference cannot be resolved here by EBSD. As expected, **Fig. 7** confirms that data points within clusters possess similar crystal orientations within a rather small range: A small variation of predominant colors denoting the crystal direction aligned with SD can easily be identified for each cluster (e.g. blue tones – close to the <111> corner of the fundamental region for CR3).

**Figs. S6** and **S7** present an analogous analysis for experiments 2 and 3. There are some areas with an above / below average proportion of data points: For example, fewer CR1 data points appear to be located in the left half of the right grain in experiment 2 (**Fig. S6b**). In experiment 3, CR2 and CR3 data points appear to be somewhat concentrated in the upper half of the wear track (**Fig. S7c/d**). As no systematic variation across experiments is apparent, this seems of limited relevance and could for example be related to the interplay of wear track topography (cf. **Fig. 6**) and EBSD backscatter geometry rather than constituting an important mechanism.

Another pivotal observation in **Figs. 7/S6/S7** is that data points in all clusters, grains and experiments are arranged in localized horizontal rows. This resembles the pattern typically seen in SEM images taken of tribologically loaded surfaces, exhibiting signs of mild ploughing. A similar surface pattern can also be seen in the SEM images of the wear tracks of experiments 1 and 2 depicted in **Fig. 9** (which will be discussed in detail later). As has been pointed out, type 2 lattice rotations (CL2 / CR2) are generally associated with larger rotational angles than type 1 (CL1 / CR1). Additionally, rotational angles exhibit a bimodal distribution in the right grain in experiments 1 and 3 (CR1 v. CR2). The present observation of a spatial arrangement of clusters in row-like features may help rationalize why this is the case: During the so called running-in of tribological systems, surface asperities are commonly smoothened. This changes surface roughness, topography and conformity and influences friction (cf. e.g. [59]). We hypothesize that such smoothing of copper asperities (inherent to the initial surface roughness) as well as interactions with the sapphire sphere's surface roughness lead to traces of locally different plastic strain. Such differences in strain could be associated with the differences in rotational angles between type 1 and 2 lattice rotations, but may at the same time influence the ability to index and evaluate the respective data points by EBSD.

As already pointed out, type 2 clusters systematically possess a lower mean band contrast than type 1 clusters (cf. **Table 2**), which would in principle be compatible with higher plastic strain. EBSD image quality metrics like band contrast, pattern quality or image quality are however prone to a number of influences and uncertainties, such as plastic and elastic strain or surface topography (cf. e.g. [60,61]). **Figs. S8** and **S9** presents maps of the band contrast per cluster for experiment 1 and 2, similar to the depiction of crystal orientation in **Figs. 7** and **S6**. For both experiments, areas of above average BC can be identified (see white arrows in **Figs. S8** and **S9**) that coincide with areas of crystal orientation that stand out in the orientation maps of the respective cluster in **Figs. 7** and **S6**, indicating a possible influence of the degree of crystal rotation on band contrast (and vice-versa). At the same time, BC values deviate markedly between different areas of individual clusters despite overall similar orientations (cf. e.g. **Fig. S8d** or **S9a/b**), indicating the existence of additional influences, such as for example wear track topography.

**Fig. 8** allows further scrutiny of a possible interdependence of type and degree of crystal rotation, plastic strain and final surface roughness. **Fig. 8a/b** shows the spatial arrangement of data points per cluster for experiments 1/2, with each cluster colored in a single color, irrespective of the precise crystal orientation of each data point. Unindexed pixels are colored in white, all remaining data points in black. **Fig. 8c/d** presents the final surface roughness within the wear track at the corresponding locations. These images were obtained from the 3D profilometry measurements in **Fig. S10** by subtracting a third-order polynomial after masking only the wear track, effectively flattening the topography. This allows observing local height differences. These roughness maps were manually registered within the black frames with respect to the depiction of the clusters in **Fig. 8a/b**. Orange arrows mark locations of type 2 crystal rotation (CL2 / CR2) in **Fig. 8a/b** and were replotted at the same locations in **Fig. 8c/d**. Red arrows, in contrast, mark deep grooves (appearing dark) in the wear track surfaces in **Fig. 8c/d** and were replotted in **Fig. 8a/b**. It is evident that at a majority of the locations marked by arrows, grooves (likely associated with higher plastic strain) coincide with type 2 clusters, exhibiting greater crystal rotations than type 1 clusters. This is however not strictly the case at all locations marked by arrows, and uncertainties due to flattening and limited lateral resolution of the topography maps as well as manual image registration persist. While certainly not providing definitive proof, **Fig. 8** does nevertheless indicate and corroborate a possible interconnection of traces of locally elevated plastic strains with higher rotational angles – within the given limitations. In conclusion, the row-like spatial arrangement of type 1 and 2 crystal rotations is likely interconnected with locally different plastic strains due to the final wear track surface topography and roughness, but at the same markedly influenced by local differences in EBSD measurement conditions. **Fig. 8** also provides evidence supporting the early speculation by Heilmann et al. that rotational components around a direction parallel to SD may be related to lateral flow and surface topography [18].

It is conceivable that the differences in crystal orientation between left and right grain, or when changing the sliding direction, also have an influence on crystal rotations. This may be responsible for the differences in magnitude and distribution of rotational angles observed for the left and right grain and between experiments 1/3 and 2. One may further speculate that depending on the specific geometry of slip systems with regard to the tribological loading (as a consequence of crystal orientation and sliding direction), certain orientation configurations may require less work to rotate than others. Friction coefficients in all three experiments are higher in the left than in the right grain (cf. **Fig. 5**), indicating that more energy was dissipated while sliding on the left grain, irrespective of the sliding direction. Caution needs to be exercised if one speculates that this may indicate it is more energy-intensive for the crystal lattice in the left grain to accommodate lattice rotation. The COF is a macroscopic quantity subject to a wide variety of influences, including the volume of material displaced (cf. mean profiles in **Fig. 6**), load, contact geometry, and both surface and roughness effects. These results however corroborate that friction is both sensitive to crystallographic orientation (left v. right grain) as well as sliding direction (experiments 1/2 v. experiment 3), as may be expected, considering the well-known impact of plastic deformation on tribological systems' friction responses [2,62–65]. Additionally, tribologically induced lattice rotations have been reported to occur in conjunction with simple shear, complicating the situation further [19]. A detailed discussion is thus beyond the scope of this work, but it seems important to point out that accounting for lattice rotations around TD (possibly in conjunction with simple shear) may be a possible lever in developing tribological systems with superior tribological properties (e.g. low friction). This proposition may help spark and guide future applied research in this direction, e.g. in terms of the development of crystallographic textures exploiting this lever by easily accommodating combined rotation around TD and shear, thereby minimizing energy expended (and likely friction). As reported, mean crystal orientations observed for clusters CR1 and CR3 match up extremely well between experiments 1 and 3 (cf. **Figs. 3b** and **4d**). This indicates that crystal orientation appears to have a systematic influence on lattice rotation magnitudes (but not on rotational axes). A detailed analysis of the systematic influence of grain orientation and sliding direction on crystal lattice rotations of type 2 will be the focus of section 4.4.

Finally, one may wonder how accurately the discussed lattice rotations represent the microstructure beneath the sliding interface. In this context, the maximum information depth beneath the surface contributing to the EBSD signal is important. It is closely related to the spatial resolution and influenced by the EBSD measuring conditions (detector, acceleration voltage, probe current, SEM electron source etc.) [49,66,67]. Chen et. al report a depth of ~60 nm contributing to the signal for copper at 20 kV and 10 nA [67]. More generally, 10-40 nm have been reported for metals [49], and more recent works suggest values two or three times as large [68,69]. In conjunction, it seems reasonable to assume that the orientation information in the present measurements likely stems from the first 50-100 nm beneath the surface. This is in agreement with the good compatibility of **Figs. 4c/d** and **S5c** for experiment 3, and corroborates that local smoothing of asperities could indeed constitute an additional, superimposed effect on lattice rotations and may in part possibly be responsible for the differences between type 1 and type 2 lattice rotations: Despite the wear track profile depth being in the range of microns (cf. **Fig. 6**), the initial copper surface roughness of ~60 nm (cf. section 2.2) is in the same order of magnitude as the signal information depth. It is thus conclusive to expect to be able to measure an influence of local smoothing of asperities by EBSD despite overall plasticity extending several microns below the initial surface.

During the early stages of dry sliding on copper, it has been reported that plastic deformation and lattice rotations are strongest near the surface and often confined to the first 100-400 nm [19,41,42]. One may thus conclude that the lattice rotations reported here are likely compatible to those reported in literature for similar experiments [18,19]. The gap in data points between LI / RI and CL1 / CR1 marked in **Fig. 2b** may constitute a lower limit of lattice rotation due to dislocation self-organization processes: During tribological loading of copper, the formation of one or more linear discontinuities entitled dislocation trace line (DTL) parallel to the sliding interface has been reported [19,42,43]. As DTLs are associated with a discrete crystal rotation around TD in close proximity to the discontinuity, the existence of this gap is conclusive. One may further hypothesize that the bimodal distribution of misorientation angles in the right grain of experiments 1 and 3 (type 1, small angle / type 2, large angle) is in a way similar in nature to the occurrence of two DTLs: Each DTL is associated with a misorientation around TD, which add up from bulk to surface. The experiment 3 TKD validation measurement – while capturing only a small area in the vicinity of the GB compared to hundreds of microns in EBSD – is in principle compatible with this view: Plastic deformation (here in terms of crystal rotations) are confined to the first few hundred nanometers beneath the sliding interface in both grains, and a layer-like structure showing some resemblance to two DTLs is observed inside a large subgrain in the right bicrystal grain (cf. **Fig. S5a**).

## 4.4   Systematic influence of slip system geometry (type 2)

It was shown that inverting the sliding direction not only inverts the sign of the rotation axis around TD from +TD (experiment 1) to –TD (experiment 2) for type 1 and 2 lattice rotations, but simultaneously also that of the superimposed rotation around SD from –SD (experiment 1) to +SD (experiment 2) for type 2. This requires further consideration. First of all, it is obvious that inverting the sliding direction changes how the twelve face-centered cubic {111}<110> slip systems in both grains are oriented with respect to the complex stress field imposed on the copper lattice by the sliding sphere. This is true even if one does not make any assumptions about the stress field itself at all, other than it being asymmetrical in terms of the sliding direction [33]. Consequently, the resolved shear stresses and thus the activation of slip systems may be expected to change. In this regard, the observed anisotropy of mean wear track profile shapes with regard to the sliding direction reported for both grains (**Fig. 6a/d** versus **Fig. 6c**) may be rationalized. This observation is well compatible with the inversion of the sign of type 2 crystal rotations around ±SD.

**Fig. 9** displays secondary electron images acquired in the vicinity of the GB for experiments 1 (**Fig. 9a**) and 2 (**Fig. 9b**). One distinct, primary set of slip traces is visible on both sides of the wear track for both experiments and in both grains, as illustrated by the four white lines in **Figs. 9a/b**. The angles between these primary sets of slip traces and SD are reported in **Table S2**. They are well compatible with the orientation of slip traces obtained by calculating the traces of the four {111} planes in the SD-TD plane using the crystal orientations measured outside of the wear track per grain (LO-mean / RO-mean). As **Table S2** demonstrates, the deviation between the angles determined by SEM and the corresponding angles calculated from EBSD are between 0.1° and 2.2°, considering the best fitting {111} plane in each case. It can be seen that when inverting the sliding direction, the primary active slip plane changes at all four locations (it is unknown which slip system on each plane produced the slip traces). In fact, with the exception of the right edge of the wear track in the right grain in experiment 2 (when facing along SD, see lower right corner of **Fig. 9b**), active slip planes interchange sides from the left to the right side of the wear track (and vice versa) when reversing the sliding direction. Discussing the interdependence of stress state, profile shapes (**Fig. 6**) and specific slip systems / slip traces at either sides of the wear track in detail is beyond the purpose of this work. Nevertheless, we may conclude that the stress state at the upper and lower edges of the wear tracks must have a significant shear stress component of opposite sign in the direction of TD, together with an obvious shear stress component in SD. The changes in active slip plane allows formulating the following hypothesis: a model slip system oriented arbitrarily will see an inversion of the sign of the resolved shear stress as a consequence of changing the sliding direction and the position within the wear track, and will respond by altering the sign of dislocations accommodating slip on this slip system. This purely geometric assumption is in agreement with the observation that a change in sliding direction results not only in an inversion of the sense of rotation around ±TD, but also that of the superimposed rotation around ±SD for type 2 lattice rotations. It is furthermore even compatible with a more macroscopic, idealized picture that due to (crystal) symmetry, an imaginary observer positioned next to the wear track in one of the grains would have to see the same results following an inversion of sliding direction (rotation around ND by 180°) when simultaneously changing the observer's position from one side of the wear track to the other. In an idealized case, this applies to the crystal rotations observed as well as the wear track shape and the activation of slip planes. One may thus conclude that the orientation of slip systems with respect to the tribological loading does in fact exert a systematic influence on how lattice rotations are accommodated – despite the fundamental nature of lattice rotation kinematics being retained irrespective of crystal orientation and load. This is further corroborated by the fact that type 3 rotations and twinning – especially sensitive to the sense of shear [57] – are only observed for sliding in +SD and in the right grain (cf. section 4.1). These insights may be fundamental in guiding future computational modelling efforts, aiming at unravelling the influence of the orientation of individual slip systems on tribologically induced crystal rotations.

## 5   Conclusions

The present work demonstrates the feasibility of large-area EBSD measurements for the investigation of early-stage crystal lattice rotations induced by dry sliding on copper. This approach offers statistics superior to smaller-scale techniques like cross-sectional EBSD or methods requiring electron transparent samples. At the same time, the risk of introducing artifacts is obviated by omitting any extra sample preparation steps before analysis. The following conclusions ensue:

- Crystal lattice rotations are categorized into three fundamental types. For a majority (50-80%) of indexed data points, crystal rotations agree well with a rotation of roughly up to 35° around the transverse direction TD (type 1). Around 10-30% exhibit larger rotation angles around TD, superimposed by a rotation around the sliding direction SD (type 2). A small fraction of indexed data points (2%) in the right grain exhibits type 1 rotation in conjunction with deformation twinning (when not changing the sliding direction).

- Inverting the sliding direction inverts the sense of the rotation around TD (type 1) and around both TD and SD (type 2) in both grains, but not the fundamental crystal rotation kinematics. This is rationalized in terms of geometrical considerations and compatible to anisotropic wear track profiles, coefficients of friction and slip trace patterns in the left and right grain.

- Increasing the load decreases EBSD data quality and increases type 1 crystal rotation prevalence over type 2. It does not otherwise fundamentally impact lattice rotation kinematics.

- Employing a large bicrystal facilitates superior EBSD measurement statistics per grain. This allows the study of multiple grain orientations while precluding the influence of the grain boundary's presence on the assessment of lattice rotation kinematics. It is demonstrated that the grain boundary does not measurably impact crystal lattice rotation kinematics in terms of the present results and conclusions, despite its impact on its immediate vicinity of a few microns.

- Lattice rotations around ±TD are at the heart of tribologically induced lattice rotation kinematics as a result of the combination of normal and shear load at the partially constrained sliding interface, irrespective of grain orientation, sliding direction and load. Slip system geometry exerts a systematic influence on how crystal rotations are accommodated, especially for data points exhibiting large rotational angles (type 2).

- Factoring in lattice rotations around the transverse direction may constitute a possible lever to be exploited in more applied future research geared towards tailoring materials and systems with superior tribological properties. Such may be achieved for instance by developing crystallographic textures especially susceptible or resistant to combined lattice rotations and shear.

## Acknowledgements

Partial funding for this work was provided by the European Research Council under ERC Grant Agreement No. 771237, TriboKey. A CC BY license is applied to the postprint of this article arising from this submission, in accordance with the grant's open access conditions. We thank R. Kilian and R. Hielscher for helpful suggestions and discussions related to MTEX.

## Competing Interests

The authors declare no conflict of interest.

**Figures** [All figures should appear in color online and in black and white in print]

**Figure 1: Experimental setup and bicrystal grain orientations. (a):** Schematic illustration of experiments 1 – 3 on copper bicrystal. The sliding direction was inverted in experiment 2. Experiment 3 was carried out with a quadruple normal load of 8 N as opposed to 2 N in experiments 1 and 2. The length of each single trace, unidirectional wear track is 10 mm. **(b):** (111) pole figure of left (mean L, yellow) and right (mean R, orange) grain mean orientations determined by EBSD in the vicinity of the grain boundary next to the wear track of experiment 1. **(c):** (100) pole figure analogous to **b**.

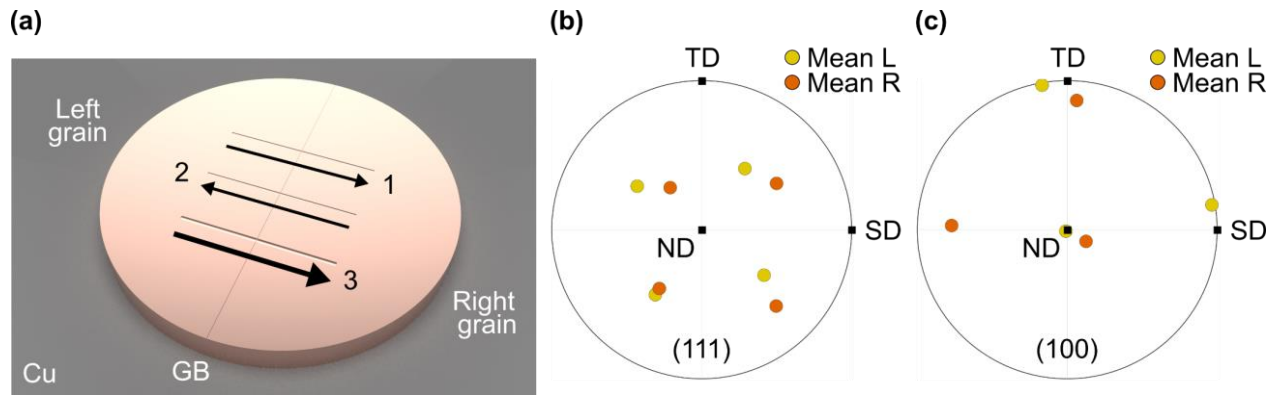

**Figure 2: Crystal orientations measured by EBSD after experiments 1 – 3. (a/c/e):** EBSD maps for experiments 1 – 3 in the vicinity of the grain boundary. The inverse pole figure color code refers to the sliding direction (SD). White pixels depict non-indexed data points. Each map is comprised of four regions: LI (left inside) and RI (right inside) inside the wear track and LO (left outside) and RO (right outside) outside of the wear track in the left and right grain, respectively. LO and RO both comprise two segments, one above and one below the wear track. The arrows above the maps indicate the sphere's sliding direction (in SD for experiments 1/3, in –SD for experiment 2). **(b/d/f):** (111) pole figures corresponding to EBSD data of experiments 1 – 3 (cf. **a/c/e**). Each EBSD data point is colored according to region it belongs to: LO and RO in yellow and orange, LI and RI in green and blue.

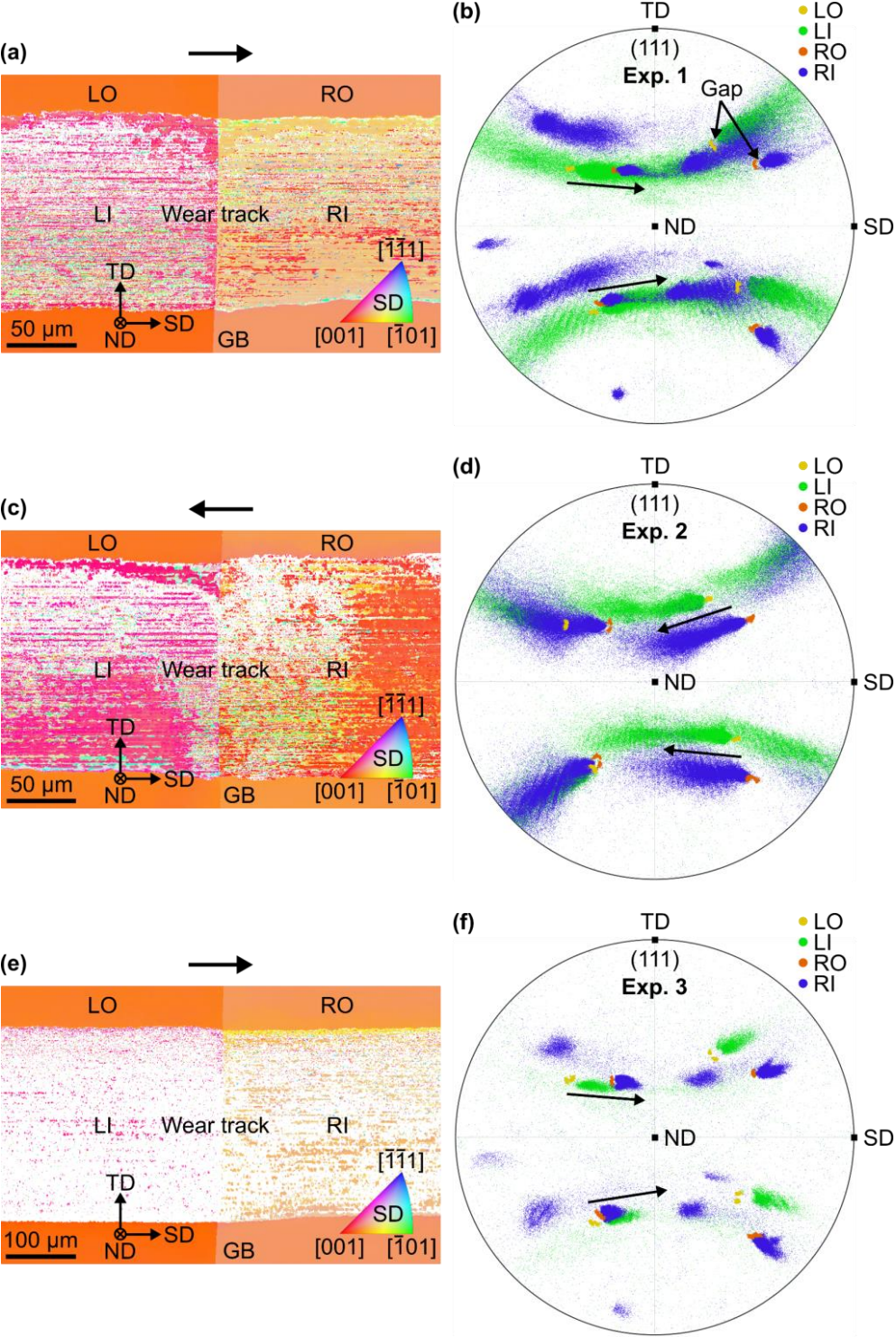

**Figure 3: Experiment 1 pole figures (PF) and inverse pole figures (IPF) for orientation data segmented into clusters.** Dotted red lines and crosses (CL sel. / CR sel., sel. = selection) in the PFs delineate the polygons used for cluster definition. Black and purple lines in **a** and **b** visualize the rotation fiber describing a perfect rotation of the copper lattice around the transverse direction (TD) by an arbitrary angle, starting from the left and right grains' initial mean orientation LO-mean (yellow, **a**) and RO-mean (orange, **b**). Pairs of black and purple arrows (one / two stars) in **a** and **b** illustrate the traversal of data points across hemispheres along the fibers. The EBSD data per cluster (CL1-CL2, cluster left 1-2 in the left grain / CR1-CR3, cluster right 1-3 in the right grain) is plotted on top of all data in regions LI (green) and RI (blue). Arrows summarize the primary direction of rotation of the corresponding clusters. **(a):** (111) PF, left grain. **(b):** (111) PF, right grain. **(c):** IPF in normal direction (ND), left grain. **(d):** IPF in ND, right grain.

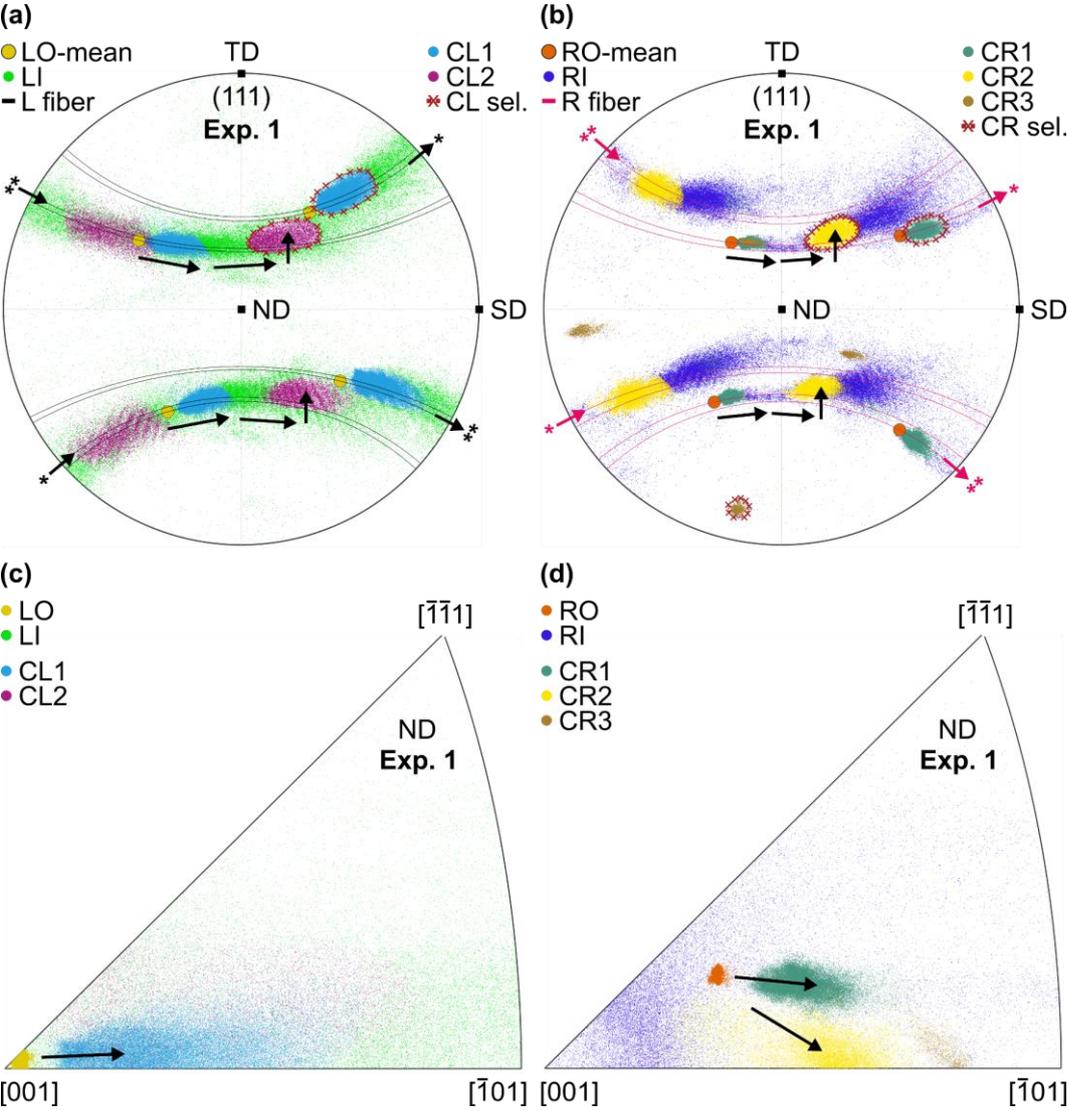

**Figure 4: Experiment 2 (inverse sliding direction, in –SD) and 3 (quadruple normal load) pole figures (PF) for orientation data segmented into clusters.** The figure layout and color scheme is identical to that described in the caption of **Fig. 3**. The lattice rotations observed for experiment 2 (**a/b**) are opposite to those of experiment 1 (cf. **Fig. 3**), those in experiment 3 (**c/d**) congenial to experiment 1. **(a):** Experiment 2 (111) PF, left grain. **(b):** Experiment 2 (111) PF, right grain. **(c):** Experiment 3 (111) PF, left grain. **(d):** Experiment 3 (111) PF, right grain.

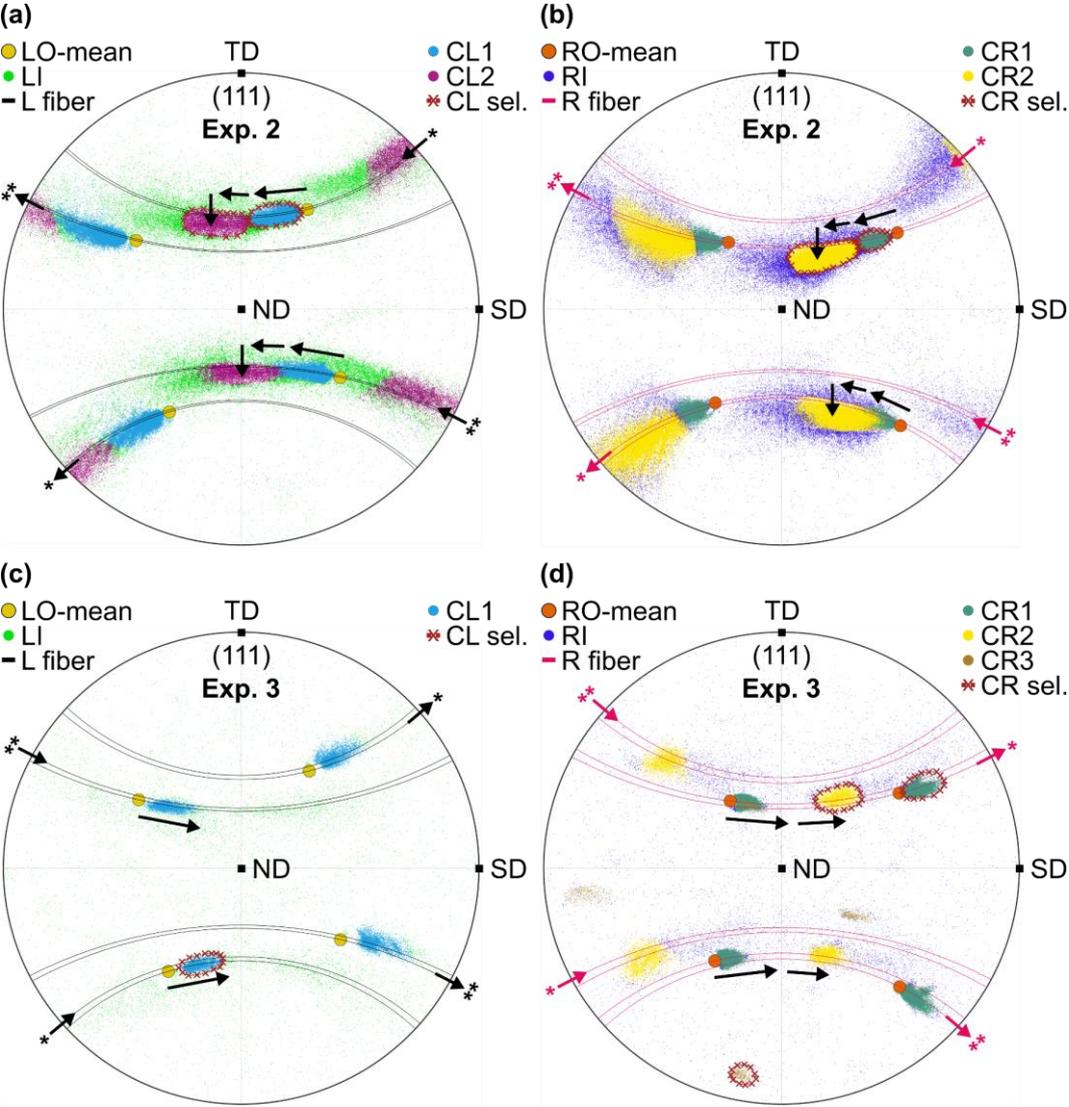

**Figure 5: Coefficient of friction (COF) during experiments 1 – 3 plotted as a function of sliding distance.** The grain boundary is traversed after approximately 5 mm of sliding in each experiment (vertical line). **(a):** COF during experiments 1 (black) and 3 (quadruple normal load, purple). Sliding from left to right grain. A step in COF is visible at the GB, friction is higher in the left than in the right grain. **(b):** COF during experiment 2. Sliding from right to left grain due to inverse sliding direction.

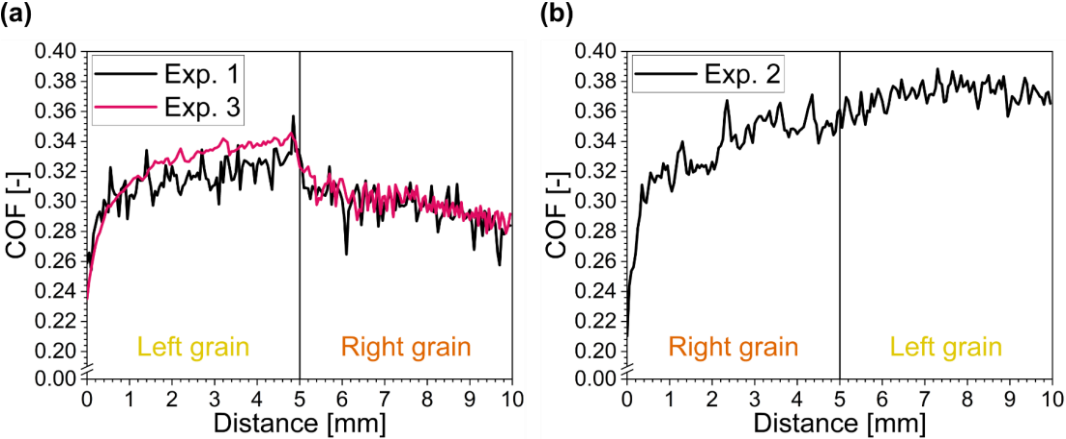

**Figure 6: Mean wear track profiles for experiments 1 – 3 in the left (L, yellow) and right (R, orange) grains as determined in the vicinity of the grain boundary (GB) by optical profilometry.** The mean profiles of experiment 1 in **a** were computed from the height map in **b** by extracting a profile in the ND-TD plane at every pixel in SD direction after leveling the measurement and averaging all profiles left / right of the GB to obtain the mean profiles L / R. Mean profiles in **c** and **d** for experiments 2 and 3 were obtained in the same fashion from corresponding measurements. The pile-up height at the left and right sides of the wear track is anisotropic with respect to the left and right grains. Relative pile-up heights in experiment 2 are qualitatively inverse to those observed for experiments 1 and 2.

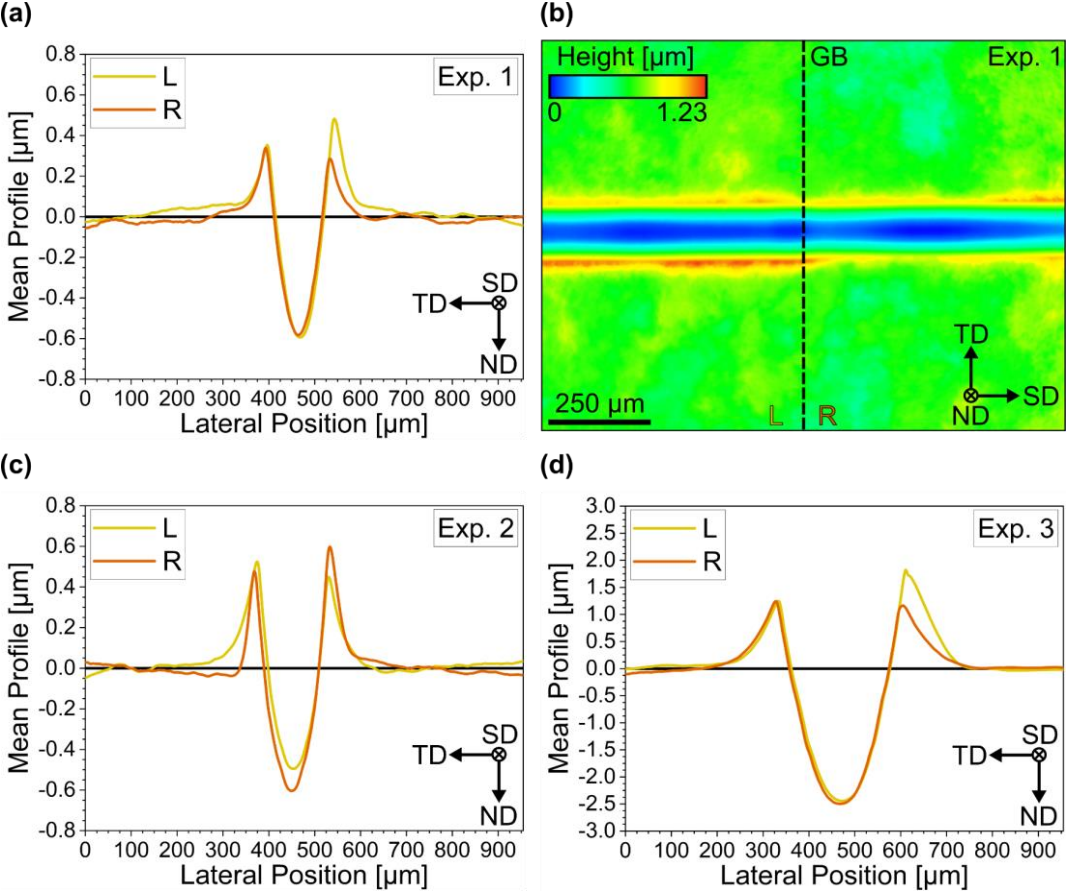

**Figure 7: Spatial distribution of EBSD data points inside clusters for experiment 1.** Each map plot only contains the data points inside the respective cluster, colored with respect to the sliding direction SD. All other data points (including non-indexed pixels) were omitted from the plot. The white lines represent the circumference of the wear track and the grain boundary (outside of the wear track). The horizontal arrows indicate the sphere's sliding direction (in SD). The left column (**a/c**) depicts clusters CL1 – 2 inside the left grain (LI), the right column (**b/d/e**) those (CR1 – 3) inside the right grain (RI). This visualization allows appraising the spatial distribution of data points inside wear track, exhibiting lattice rotations characteristic for the respective cluster. **Figures S6** and **S7** in the supplementary information display the same information for experiments 2 (inverse sliding direction) and 3 (quadruple normal load).

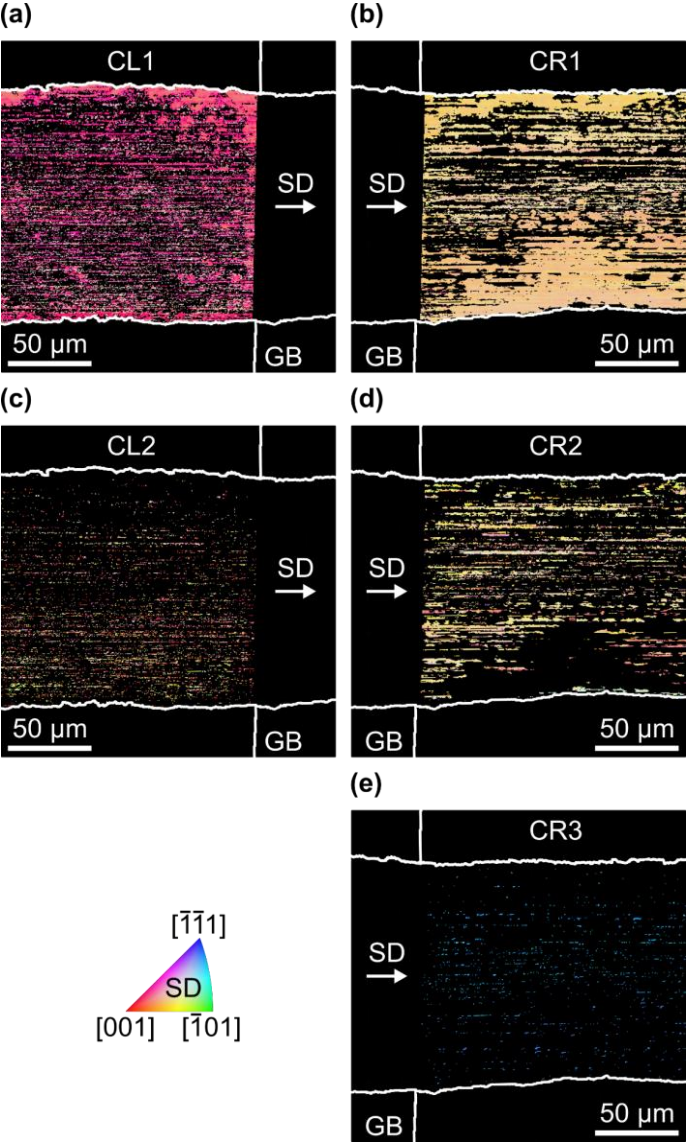

**Figure 8: Comparison of spatial distribution of clusters & surface roughness inside the wear track in the vicinity of the grain boundary (GB). a** and **b** show the spatial distribution of clusters in both grains for experiments 1 and 2. All data points belonging to each cluster are colored in a single color, irrespective of orientation. Unindexed data points appear in white, all remaining data points in black. Images **c** and **d** depict the corresponding surface roughness inside the wear track and were manually registered inside the black frame with regard to **a/b**. Orange arrows mark areas belonging to type 2 lattice rotations (clusters CL2 / CR2) in **a/b** and were replotted at the same locations in **c/d**. In contrast, red arrows mark deep groves (appearing dark) within the wear track in **c/d** and were replotted in **a/b**. Despite the uncertainty due to manual image registration, a trend of type 2 rotations to coincide with deep grooves and vice-versa can be observed at many, but not all locations marked by arrows.

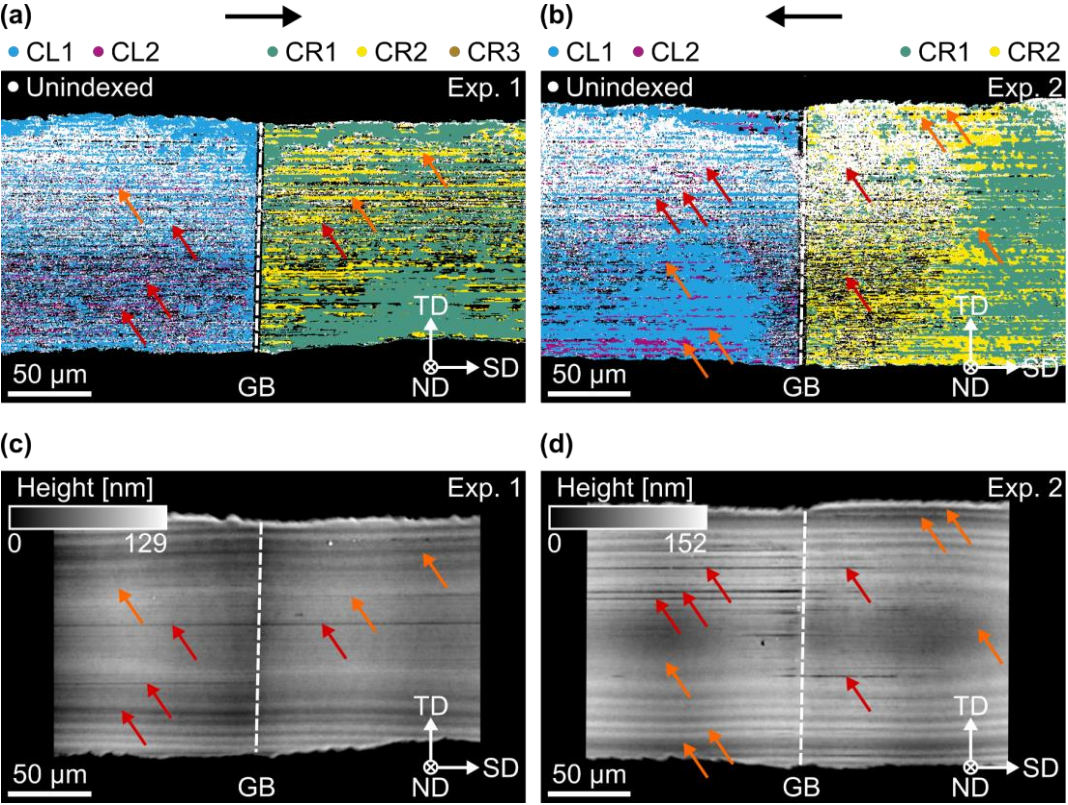

**Figure 9: Slip traces in the vicinity of the grain boundary (GB) for experiments 1 (a) and 2 (b, inverse sliding direction, see arrows) revealed by secondary electron images.** White lines visualize the trace of the most active slip plane (as deduced from the images) on the left and right sides of the wear track for the left and right grains (L / R). It is evident that changing the sliding direction changes the primary active slip plane at all four locations (both grains, both sides of the wear track). Note that it is not directly possible to assess which of the three slip systems belonging to each slip plane is active.

**(a)**

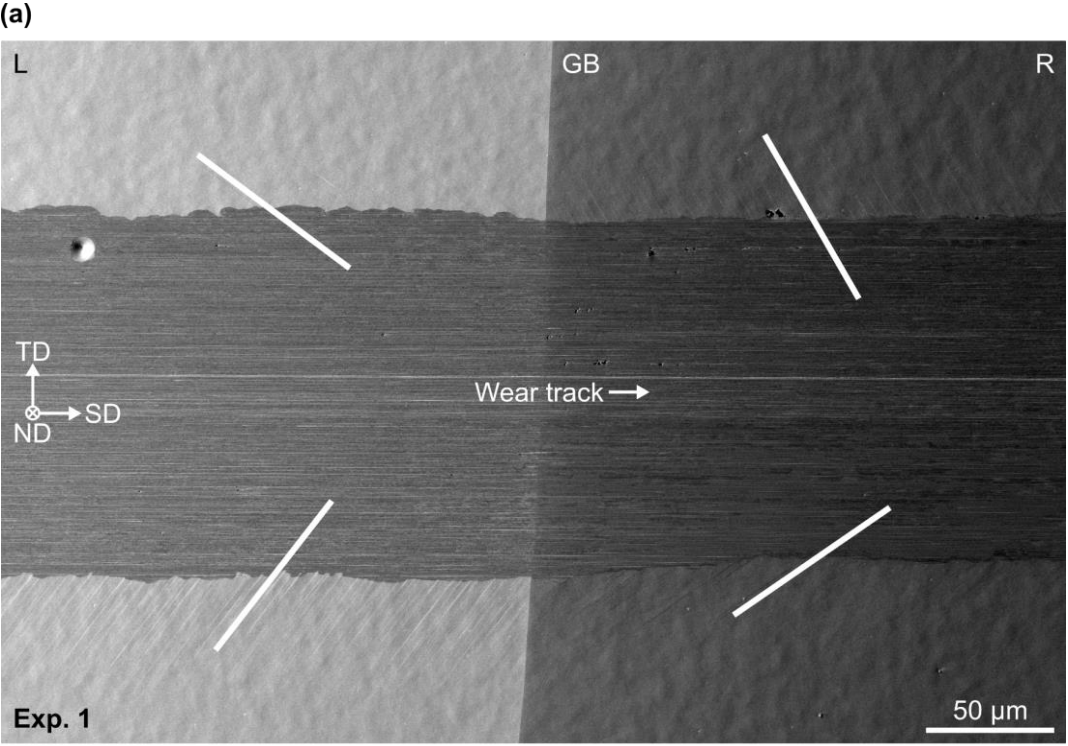

**(b)**

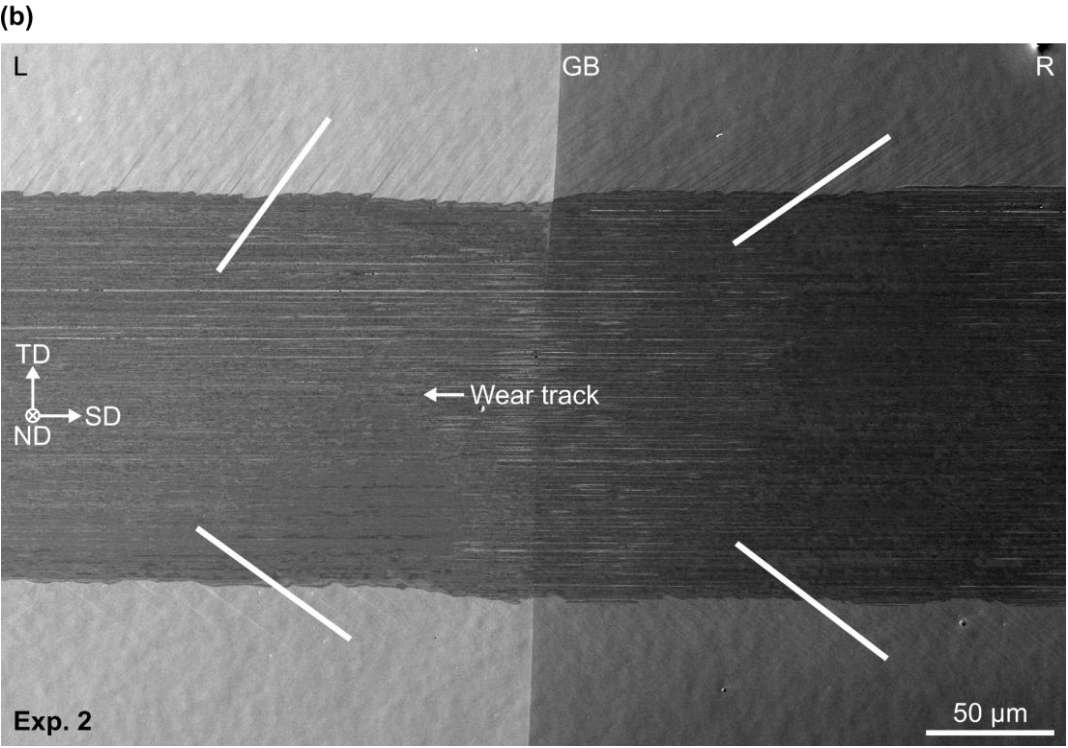

# Tables

**Table 1: EBSD indexing and data quality overview.** For each experiment, information is provided on the two regions outside (LO and RO) and inside the wear track (LI and RI), respectively. The total number of data points as well as the proportion of indexed data points thereof is given (note that regions LO / RO are by definition only comprised of indexed pixels). Data quality per region is provided in terms of mean band contrast (BC) and the mean number of indexed Kikuchi bands (of a maximum of twelve for face-centered cubic crystal symmetry).

| experiment | region | no. px. [-] | indexed [%] | band contrast (BC) [-] | | no. indexed bands [-] | |
|:---:|:---:|:---:|:---:|:---:|:---:|:---:|:---:|
| | | | | mean | std. dev. | mean | std. dev. |
| 1 | LO | 57321 | 100 | 207 | 6 | 11.1 | 0.6 |
| | RO | 66608 | 100 | 199 | 7 | 11.7 | 0.5 |
| | LI | 108146 | 66 | 150 | 27 | 8.5 | 1.9 |
| | RI | 105990 | 90 | 176 | 26 | 9.8 | 1.8 |
| 2 | LO | 48511 | 100 | 196 | 7 | 10.2 | 0.6 |
| | RO | 43462 | 100 | 210 | 14 | 10.8 | 0.8 |
| | LI | 119831 | 69 | 171 | 32 | 8.8 | 1.8 |
| | RI | 125095 | 73 | 173 | 30 | 8.8 | 1.8 |
| 3 | LO | 59482 | 100 | 193 | 9 | 10.7 | 0.7 |
| | RO | 63952 | 100 | 206 | 12 | 11.9 | 0.4 |
| | LI | 108249 | 9 | 116 | 25 | 7.2 | 1.5 |
| | RI | 105046 | 30 | 133 | 28 | 8.1 | 1.8 |

**Table 2: Quantification of proportions of number of data points and data quality per cluster.** Information on the number of indexed pixels per cluster (IP) as a proportion of the number of pixels in the parent region (PR, see below). Band contrast and the mean number of indexed bands as measures for data quality are provided for all clusters per grain and experiment (cf. **Figs. 3/4** for definition of clusters). Each cluster is a subset of all data inside the wear track for the corresponding experiment and grain, its so called parent region. PI (proportion of *indexed* pixels) states how many percent each cluster's indexed pixels make up of all *indexed* pixels in its parent region (PR). PT (proportion of *total* pixels) is the same metric, but with respect to *all* pixels in the parent region (i.e. including all non-indexed pixels in the parent region). Note that a cluster – other than regions LI / RI – is by definition only comprised of indexed pixels.

| experiment | parent region (PR) | cluster | Indexed pixels (IP) in cluster as proportion of | | band contrast (BC) [-] | | no. indexed bands [-] | |
|---|---|---|---|---|---|---|---|---|
| | | | PR indexed px. (PI) [%] | PR total px. (PT) [%] | mean | std. dev | mean | std. dev. |
| 1 | LI | CL1 | 60 | 39 | 156 | 28 | 8.9 | 1.9 |
| | | CL2 | 12 | 8 | 142 | 22 | 7.9 | 1.6 |
| | RI | CR1 | 58 | 52 | 186 | 19 | 10.4 | 1.5 |
| | | CR2 | 19 | 17 | 170 | 29 | 9.5 | 1.9 |
| | | CR3 | 2 | 2 | 176 | 26 | 8.8 | 1.4 |
| 2 | LI | CL1 | 72 | 49 | 177 | 30 | 9.1 | 1.7 |
| | | CL2 | 12 | 8 | 164 | 33 | 8.2 | 1.7 |
| | RI | CR1 | 52 | 38 | 187 | 26 | 9.4 | 1.6 |
| | | CR2 | 31 | 23 | 163 | 27 | 8.4 | 1.8 |
| 3 | LI | CL1 | 71 | 6 | 121 | 25 | 7.5 | 1.6 |
| | RI | CR1 | 78 | 24 | 135 | 28 | 8.3 | 1.8 |
| | | CR2 | 12 | 4 | 129 | 25 | 7.8 | 1.7 |
| | | CR3 | 2 | 0 | 120 | 23 | 7.6 | 1.4 |